\documentstyle[amstex,amssymb,preprint,aps]{revtex}
\tightenlines

\begin{document}
\title{The Spin}
\author{S. Danko Bosanac}
\address{R. Bo\v{s}kovi\'{c} Institute, 10001 Zagreb \\
Croatia}
\maketitle

\begin{abstract}
It is shown that the spin is naturally introduced into classical mechanics
if the latter is formulated as dynamics of the phase space density. It is
shown that the uncertainty principle, as the amendment in this dynamics,
restricts possible spins, and in particular equation for the particle with
the spin $\hbar /2$ is derived. Also equation for the charge with this spin
is derived when electromagnetic field is included. In one example it is
shown that the modulus of the spin changes with the gradient of the magnetic
field.
\end{abstract}

\section{Introduction}

There are a number of parameters that uniquely specify a particle, and one
of them is the spin\cite{tomonaga}. Yet there is a fundamental difference
between, say the charge and the mass of a particle and its spin. The former
are arbitrary parameters that enter dynamics equations, but the spin is the
property of the dynamics equation, the Dirac equation\cite{dirac}.
Therefore, in principle its explanation should be traced to the principles
from which the Dirac equation is derived, and in this respect the spin
appears not as a fundamental property. Despite this observation the
explanation for the spin withstood various attempts, so that nowadays it is
considered a property of the particle that does not have classical
description. Nevertheless it is worth mentioning one attempt to explain it
as a classical effect\cite{mcgregor}. For the electron to have the spin it
is assumed to be a finite sized sphere (all experimental evidence shows the
contrary) with certain distribution of the charge on its surface or the
volume (what keeps the charge together is avoided as an answer) that rapidly
rotates (to get the value $\hbar /2$ for the spin superluminal equatorial
velocity of the sphere must be assumed). The inability of the classical
model (it is almost impossible to think of another classical model for the
spin\cite{berezin}) to explain the spin resulted in the generally accepted
view that in the quantum world there are phenomena that cannot be explained
by our experience, and one of them is the spin\cite{bohr,mott}. Therefore
one does not make models of the spin, one uses the rules by which it is
mathematically described. One should say, though, that there are various
semiclassical studies of the dynamics of the charges with the spin, which
can be called empirical rules. In the classical equations of motion one
incorporates the term that represents magnetic dipole\cite
{bargmann,barut,batelaan,garraway} (which is point-like and manifestation of
the spin) interaction with the magnetic field, without questioning the
source of this dipole.

In this work it will be shown that the spin has natural description from the
classical principles, but it has the feature that its value is arbitrary.
However, by assuming the amendment that classical principles should be
consistent with the uncertainty principle the spin is fixed and has the
value $\hbar /2$. This assumption in classical mechanics has been elaborated
and tested in a number of cases\cite{bos1,bos2,bos3,bos4}, and therefore the
idea behind it will only be briefly described here. One starts from the
observation that the deterministic view of classical mechanics, which is
based on the idea that given initial position of the particle in the phase
space its future and past is entirely determined if the forces on it are
known, is fundamentally wrong. Even in principle one cannot imagine making
experiment that does not incorporate inaccuracies in the measurement of the
position of particles in the phase space, and this means that one only knows
probability of their whereabouts. If one accepts this view then the
particles in the phase space are not represented by points, but by phase
space (probability) densities, and their dynamics is governed by the
Liouville equation. Implementing uncertainty principle makes a restriction
on the possible phase space densities, which has significant impact on its
dynamics. As it was shown, this implementation results in deriving quantum
dynamics equation, the Schroedinger equation\cite{bos4}. Among the various
examples treated in this way were diffraction and tunnelling effects\cite
{bos5}, for which it was shown to be described by solving classical
equations of motion. It will be now shown how the spin is described, what is
its classical description and what some of its properties are. Although
combination of classical dynamics with the uncertainty principle was done in
the relativistic case\cite{bos6}, when it was shown that Dirac equation is
obtained, a number of points were missed in that derivation. The source of
the spin is not discussed in details, and in particular why one needs
relativistic theory to describe it. Also relativistic theory was described
for a free particle only, and it is imperative to include electromagnetic
(EM) field. In this work only non relativistic theory is described, and it
is shown that the spin is not the relativistic phenomenon. Furthermore
interaction with the EM field is described, and in one example it is shown
that the spin does not have universal value, but its modulus changes with
the strength of the field, more precisely it depends on the gradient of the
magnetic component of the EM field. When the homogeneous magnetic field is
restored (or when it goes to zero) the modulus of the spin gets back its
value $\hbar /2$.

The discussion starts with a simple classical model that has analogy with
the spin, in order to show that the concepts leading to its true description
are not alien in classical mechanics.

\section{Simple classical model of the spin}

The arguments against the possibility of classical description of the spin
are very convincing and yet one can show that this concept is naturally
introduced into classical mechanics. In the traditional classical treatment
a particle is represented as a point in the phase space whose dynamics is
determined by two parameters: its initial position in this space and the
force that acts on it. However, if the particle is represented as a
probability density function $\rho \left( \vec{r},\vec{p},t\right) $ in the
phase space, which takes into account all uncertainties connected with
determining its initial conditions, then its dynamics is solved from the
Liouville equation 
\begin{equation}
\frac{\partial \rho }{\partial t}\,+\,\frac{\vec{p}}{m}\cdot \nabla _{\vec{r}%
}\rho \;+\vec{F}\cdot \nabla _{\vec{p}}\rho =\;0  \label{li}
\end{equation}
The mass of particle is $m$ , its momentum is $\vec{p}$ and the force on the
particle is $\vec{F}$. The simplest example to solve from this equation is
for a particle that is not moving and it is placed in the origin of the
coordinate system. This means that at $t=0$ the value of these parameters
are defined as the averages 
\[
\int d^{3}r\;d^{3}p\;\vec{r}\;\rho _{0}\left( \vec{r},\vec{p}\right) =\int
d^{3}r\;d^{3}p\;\frac{\vec{p}}{m}\;\rho _{0}\left( \vec{r},\vec{p}\right) =0 
\]
and if the phase space density is a function of the module $r$ and $p$ then
the (average) angular momentum of the particle is zero. However the
(average) angular momentum squared is not zero, which is explicitly checked
for the phase space density 
\begin{equation}
\rho _{0}\left( \vec{r},\vec{p}\right) =\frac{1}{a^{3}b^{3}\pi ^{3}}%
e^{-r^{2}/a^{2}-p^{2}/b^{2}}  \label{clro0}
\end{equation}
when this quantity has the value 
\begin{equation}
<L^{2}>=\int d^{3}r\;d^{3}p\;\left( \vec{r}\times \vec{p}\right) ^{2}\;\rho
_{0}\left( \vec{r},\vec{p}\right) =\frac{3}{2}a^{2}b^{2}  \label{cll2}
\end{equation}
Therefore particle at rest in the origin of the coordinate system manifests
angular momentum, whose average value is zero, but not because its modulus
is zero but because one sense of rotation in the phase space density is
equally probable as the other. Indeed if the average over one sense of
rotation, say the positive and parallel to the x-y plane, is calculated then
the resulting (average) angular momentum is 
\begin{gather}
<\vec{L}>=\int_{>}d^{3}r\;d^{3}p\;\vec{r}\times \vec{p}\;\rho _{0}\left( 
\vec{r},\vec{p}\right)  \label{angm} \\
=\hat{z}\;\int d^{3}r\;\int_{-\infty }^{\infty }dp_{r}\int_{-\infty
}^{\infty }dp_{\theta }\;\int_{0}^{\infty }dp_{\phi }\;rp_{\phi }\sin \theta
\;\rho _{0}\left( \vec{r},\vec{p}\right) =\frac{ab}{4}\hat{z}  \nonumber
\end{gather}
where the components of the momentum with respect to the vector $\vec{r}$
were used.

That the previous finding is not only of academic interest will be
demonstrated on the example that is quite realistic. It is assumed that a
charged particle is placed in a constant magnetic field, with the assumption
that before $t=0$ the field is zero, and then its time evolution is
followed. The problem has a straightforward classical solution except that
one step is often overlooked. When one says that a particle with the charge $%
e$ is placed in the magnetic field at $t=0$ then its equation of motion is 
\begin{equation}
m\;d_{t}^{2}\vec{r}=\frac{e\vec{v}}{c}\times \left( \nabla \times \vec{A}%
\right)  \label{eqh}
\end{equation}
where the tentative assumption is that initial position $\vec{r}_{0}$ and
velocity $\vec{v}_{0}$ are determined before the field is turned on. This
assumption is fundamentally wrong because during the infinitely short time
interval around $t=0$ the vector potential changed from its zero value to $%
\vec{A}$, which means that it had a step-like time dependence. This means
that the equation of motion for the charge is 
\[
m\;d_{t}^{2}\vec{r}=-\frac{e}{c}\partial _{t}\vec{A}+\frac{e\vec{v}}{c}%
\times \left( \nabla \times \vec{A}\right) =-\frac{e}{c}\vec{A}\;\delta (t)+%
\frac{e\vec{v}}{c}\times \left( \nabla \times \vec{A}\right) 
\]
and the velocity in a very small time interval from $t=0-\varepsilon $ to $%
t=0+\varepsilon $ becomes 
\begin{equation}
\vec{v}=\vec{v}_{0}-\frac{e}{mc}\vec{A}\left( \vec{r}_{0}\right)
+O(\varepsilon )  \label{initvh}
\end{equation}
It follows that in the magnetic field initial position for the charge is the
same as when there was no field, but the initial velocity is given by (\ref
{initvh}), or the momentum $\vec{P}_{0}=\vec{p}_{0}-\frac{e}{c}\vec{A}\left( 
\vec{r}_{0}\right) $. This observation has important consequence on the time
evolution of the phase space density. Its initial functional form, which is
determined before magnetic field is turned on, is $\rho _{0}(\vec{r}_{0},%
\vec{p}_{0})$, but its time evolution is determined as if the initial
momentum of the particle has the value $\vec{P}_{0}=\vec{p}_{0}-\frac{e}{c}%
\vec{A}\left( \vec{r}_{0}\right) $. The equation of motion (\ref{eqh}) is
therefore solved for the initial $\vec{P}_{0}$, and solution for the
position is $\vec{r}=\vec{f}(\vec{r}_{0},\vec{P}_{0},t)$ while for the
momentum is $\vec{P}=\vec{g}(\vec{r}_{0},\vec{P}_{0},t)$. The phase space
density is therefore a function of the form $\rho (\vec{r},\vec{P},t)$ and
in terms of the initial phase space density it is given by 
\begin{equation}
\rho (\vec{r},\vec{P},t)=\rho _{0}\left[ \vec{f}(\vec{r},\vec{P},-t),\vec{g}(%
\vec{r},\vec{P},-t)+\frac{e}{c}\vec{A}\left( \vec{f}(\vec{r},\vec{P}%
,-t)\right) \right]  \label{roh}
\end{equation}
Angular momentum of the particle is (\ref{angm}) but $\vec{p}$ is replaced
by $\vec{P}$.

The simplest example is when magnetic field is uniform of the strength $%
h_{0} $ and it is directed along the z axes. The vector potential from which
it is derived is $\vec{A}=\frac{h_{0}}{2}\;\hat{z}\times \vec{r}$, and
solution of the equations of motion (\ref{eqh}) is (momentum $\vec{P}_{0}$
has only x and y components) 
\begin{eqnarray}
\vec{r} &=&\vec{r}_{0}+\frac{c\vec{P}_{0}}{eh_{0}}\sin \left( \frac{eh_{0}t}{%
mc}\right) +\frac{c\vec{P}_{0}\times \hat{z}}{eh_{0}}\;\left[ 1-\cos \left( 
\frac{eh_{0}t}{mc}\right) \right] +\frac{P_{0z}}{m}\;t\;\hat{z}\equiv \vec{f}%
(\vec{r}_{0},\vec{P}_{0},t)  \label{trajh} \\
\vec{P} &=&\vec{P}_{0}\cos \left( \frac{eh_{0}t}{mc}\right) +\vec{P}%
_{0}\times \hat{z}\;\sin \left( \frac{eh_{0}t}{mc}\right) +P_{0z}\;\hat{z}%
\equiv \vec{g}(\vec{r}_{0},\vec{P}_{0},t)  \nonumber
\end{eqnarray}
From these trajectories one calculates time evolution of the phase space
density from (\ref{roh}), and by assuming that prior to turning the field on
angular momentum of the particle is zero then immediately after that it is
given by 
\begin{gather*}
\vec{L}_{0}=\int d^{3}r_{0}\;d^{3}P_{0}\;\vec{r}_{0}\times \vec{P}_{0}\;\rho
_{0}\left[ \vec{r}_{0},\vec{P}_{0}+\frac{e}{c}\vec{A}\left( \vec{r}%
_{0}\right) \right] =\int d^{3}r_{0}\;d^{3}P_{0}\;\vec{r}_{0}\times \left[ 
\vec{P}_{0}-\frac{e}{c}\vec{A}\left( \vec{r}_{0}\right) \right] \;\rho
_{0}\left( r_{0},P_{0}\right) \\
=-\frac{eh_{0}}{2c}\;\hat{z}\int d^{3}r_{0}\;d^{3}P_{0}\;\left(
x_{0}^{2}+y_{0}^{2}\right) \;\rho _{0}\left( r_{0},P_{0}\right)
\end{gather*}
Therefore the particle acquires angular momentum, and because it has origin
in the internal rotation of the phase space density and not in the motion of
its center of gravity, it can be shortly called spin.

At any later time the angular momentum is 
\[
\vec{L}=\int d^{3}r\;d^{3}P\;\vec{r}\times \vec{P}\;\rho _{0}\left[ \vec{f}(%
\vec{r},\vec{P},-t),\vec{g}(\vec{r},\vec{P},-t)+\frac{e}{c}\vec{A}\left( 
\vec{f}(\vec{r},\vec{P},-t)\right) \right] 
\]
and after the change of integration variables its final expression is 
\begin{eqnarray}
\vec{L} &=&-\hat{z}\;\frac{eh_{0}}{4c}\left[ 1+\cos \left( \frac{eh_{0}t}{mc}%
\right) \right] \int d^{3}r\;d^{3}P\;\left( x^{2}+y^{2}\right) \rho
_{0}\left( r,P\right) -  \label{angmcl} \\
&&\hat{z}\;\frac{c}{eh_{0}}\left[ 1-\cos \left( \frac{eh_{0}t}{mc}\right) %
\right] \int d^{3}r\;d^{3}P\;\left( P_{x}^{2}+P_{y}^{2}\right) \rho
_{0}\left( r,P\right) \;  \nonumber
\end{eqnarray}
It is a time dependent quantity, which is always oriented parallel to the z
axes.

The same example is calculated from quantum mechanics, in order to show that
the result is not pure academic exercise. The assumption is that the initial
conditions are the same, which means that the initial wave function $f_{0}$
is such that 
\[
\left| f_{0}\left( \vec{r}\right) \right| ^{2}=\int d^{3}p\;\rho _{0}(\vec{r}%
,\vec{p}) 
\]
The function $f_{0}\left( \vec{r}\right) $ is real because it represents
particle that is not moving. Schroedinger equation is 
\[
i\hbar \partial _{t}f=-\frac{\hbar ^{2}}{2m}\left( \nabla -\frac{ei}{\hbar c}%
\vec{A}\right) ^{2}\;f\;\; 
\]
and in the Cartesian coordinates it is (only rotationally symmetric problem
is considered, in which case linear term in vector potential can be omitted)

\[
i\partial _{t}f=-\frac{\hbar }{2m}\left[ \partial _{x}^{2}+\partial
_{y}^{2}+\partial _{z}^{2}-\frac{e^{2}h_{0}^{2}}{4\hbar ^{2}c^{2}}\left(
x^{2}+y^{2}\right) \right] f 
\]
In the z direction equation is for a free particle while in the x and y
directions it is for harmonic oscillators. Both solutions are known, and the
wave function is written as expansion 
\[
f(x,y,z,t)=\sum_{n,s}\int dk\;a_{n,s}(k)f_{n}(x)f_{s}(y)e^{ikz}e^{-i\frac{%
E_{n}}{\hbar }t-i\frac{E_{s}}{\hbar }t-i\frac{\hbar }{2m}k^{2}t} 
\]
$\allowbreak $where $f_{n}(x)$ are eigenfunctions for the harmonic
oscillator. The coefficients are determined from the initial $f_{0}$, and
because of its symmetry the summation indices run over only even integers.
In the simplest model 
\[
f_{0}=e^{-\frac{1}{d^{2}}\left( x^{2}+y^{2}+z^{2}\right) } 
\]
in which case the coefficients are 
\[
a_{n,s}(k)=b_{n}b_{s}\frac{d^{1/2}}{\pi ^{1/4}}e^{-\frac{1}{2}d^{2}k^{2}} 
\]
where 
\[
b_{n}=\frac{\left( 2eh_{0}c\hbar \right) ^{1/4}}{\left( \frac{n}{2}\right) !}%
\sqrt{\frac{d\;n!}{2^{n-1}}}\frac{\left( -2\hbar c+eh_{0}d^{2}\right) ^{n/2}%
}{\left( 2\hbar c+eh_{0}d^{2}\right) ^{(n+1)/2}}\; 
\]

From this solution one would want to calculate the (average) angular
momentum, which is defined as $\vec{L}=m\int d^{3}r\;\vec{r}\times \vec{j}$
where $\vec{j}$ is the probability current, and for a particle in the
magnetic field it is given by 
\begin{equation}
\vec{j}=\frac{\hbar }{m}%
\mathop{\rm Im}%
\left( f^{\ast }\nabla \;f\right) -\frac{e}{mc}\;\vec{A}\;f^{\ast }f
\label{ja}
\end{equation}
It can be shown that the contribution of the first term in the current is
zero, and so the angular momentum is 
\begin{equation}
\vec{L}=-\frac{1}{4ceh_{0}d^{2}}\left[ 4\hbar
^{2}c^{2}+e^{2}h_{0}^{2}d^{4}+\cos \left( \frac{eh_{0}t}{cm}\right) \left(
-4\hbar ^{2}c^{2}+e^{2}h_{0}^{2}d^{4}\right) \right] \;\hat{z}
\label{angmqv}
\end{equation}

In the phase space density (\ref{clro0}) the parameters $a$ and $b$ are
arbitrary but for legitimate comparison with quantum results they must be
specified so that 
\begin{equation}
\rho _{0}\left( r,p\right) =\left| f(\vec{r})\right| ^{2}\left| g(\vec{p}%
)\right| ^{2}=\frac{1}{(\hbar \pi )^{3}}e^{-\frac{r^{2}}{d^{2}}-\frac{%
p^{2}d^{2}}{\hbar ^{2}}}  \label{rogauss}
\end{equation}
where $g(\vec{p})$ is the function $f(\vec{r})$ in the momentum space. For
this choice of the initial phase space density the angular momentum (\ref
{angmcl}) has identical value as (\ref{angmqv}), which is shown by
evaluating simple integrals. Therefore classical and quantum dynamics of a
charge in the homogeneous magnetic field produce identical solutions.

\section{The spin}

There are two important results of the previous analysis for the following
discussion. One is that if classical dynamics is formulated from the
Liouville equation then its solution is identical to the quantum. It could
be argued that the identity is accidental, but it will be shown that this is
not the case. Second result indicates that there are two contributions to
the total angular momentum of the particle. For a completely general initial
phase space density it can be shown that the total angular momentum is given
as a sum 
\begin{equation}
\vec{J}=\vec{L}+\vec{S}  \label{ls}
\end{equation}
where the first term is angular momentum which is associated with
translation of the centre of gravity of probability density $P(\vec{r},t)$,
and it is called the orbital angular momentum. The second contribution comes
from the internal rotation of the probability density and can be called the
spin. In the previous section it was shown that it is caused by the act of
turning magnetic field on, because rotating electric field is induced. The
spin is therefore not an unusual feature of the particle, but in that
example its value depends on a number of external parameters, i.e. it is
essentially arbitrary. One result in particular is intriguing, and it is
summarized in (\ref{angm}). It was found that for a phase space density that
represents stationary particle, but with spherically symmetric spatial and
momentum probability densities, the average modulus of the angular momentum
is not zero, but the angular momentum itself is. The question is can a phase
space density be found with the same symmetry properties for which the
angular momentum is not zero? Indeed it can be found, for example if one
writes 
\begin{equation}
\rho (\vec{r},\vec{p})=\rho _{0}(r,p)+\frac{1}{2}\left( \nabla _{r}\times
\nabla _{p}\right) \cdot \vec{s}\;\rho _{0}(r,p)  \label{rospin}
\end{equation}
then $<\vec{L}>=\vec{s}$ but $<L^{2}>$ is not affected by this additional
term. However, this phase space density is arbitrary because there is
nothing that prevents choosing arbitrary value for the spin $\vec{s}$ .

The identity of the classical and quantum dynamics in the previous section
is not accidental, but a consequence of selecting phase space density that
is in accord with the uncertainty principle. The principle states that if
the phase space density produces standard deviation $\Delta x$ for the
Cartesian coordinate $x$, and standard deviation $\Delta p_{x}$ for the
momentum component $p_{x}$ then at each instant in time the inequality 
\[
\Delta x\;\Delta p_{x}\geqslant c 
\]
should be satisfied. Discussion how the constant $c$ is determined from
experiment is omitted, but anticipating value is $c=\hbar /2$. The phase
space density with this property is obtained by parameterisation (in the
Fourier analysis it is called convolution, but it is also known as the
Wigner function\cite{wigner,hillery,carruthers,moyal}) 
\begin{equation}
\rho (\vec{r},\vec{p}.t)\;=\;\frac{1}{\hbar ^{3}\pi ^{3}}\int d^{3}q\,e^{2i%
\vec{p}\cdot \vec{q}/\hbar }f^{\ast }(\vec{r}+\vec{q},t)f(\vec{r}-\vec{q},t)
\label{roun}
\end{equation}
where $f(\vec{r},t)$ should be determined by requiring that $\rho (\vec{r},%
\vec{p}.t)$ satisfies the Liouville equation (\ref{li}). For a free particle
this substitution results in the equation 
\[
\int d^{3}q\,e^{2i\vec{p}\cdot \vec{q}/\hbar }\left[ 
\begin{array}{c}
f^{\ast }(\vec{r}+\vec{q},t)\;\partial _{t}f(\vec{r}-\vec{q},t)+\frac{\hbar 
}{2im}f^{\ast }(\vec{r}+\vec{q},t)\Delta f(\vec{r}-\vec{q},t)+ \\ 
f(\vec{r}-\vec{q},t)\;\partial _{t}f^{\ast }(\vec{r}+\vec{q},t)-\frac{\hbar 
}{2im}f(\vec{r}-\vec{q},t)\Delta f^{\ast }(\vec{r}+\vec{q},t)
\end{array}
\right] =0 
\]
which is satisfied if 
\begin{equation}
\partial _{t}f(\vec{r},t)+\frac{\hbar }{2im}\Delta f(\vec{r},t)=0
\label{schrfr}
\end{equation}
In the equation for the function $f(\vec{r},t)$ one recognizes Schroedinger
equation for a free particle. It can be shown, relatively easily for the
harmonic type potentials, \ that the same identity holds for a general
potential\cite{bos4}. Therefore the identity in the previous section is not
accidental but result of imposing uncertainty principle on the phase space
density. The only accident is that for the Gaussian type probabilities the
parameterisation (\ref{rogauss}) is identical to (\ref{roun}).

There is one very important consequence when the uncertainty principle is
imposed on the phase space density. If the (wave) function $f$ is assumed to
be spherically symmetric, i.e. it depends on the modulus of $\vec{r}$, then
it can be shown that no-matter what its functional form the value (\ref{cll2}%
) for the angular momentum squared is always $<L^{2}>=\frac{3}{2}$. Likewise
it can be shown that although the angular momentum (\ref{angm}) is zero its
value from only one half of the phase space is always $<\vec{L}>=\frac{1}{4}%
\hat{z}$ (from now on $\hbar $ is set to unity). These are like universal
numbers, in which case one can think of possibility to have phase space
densities whose spin is not zero, always having certain universal value. To
find the phase space density with that property one starts by calculating
probability current from (\ref{roun}), and by writing momentum as 
\begin{equation}
\vec{p}=\hat{S}\left( \hat{S}\cdot \vec{p}\right) +\hat{S}\times \left( \vec{%
p}\times \hat{S}\right)  \label{pdec1}
\end{equation}
it is given by 
\[
\vec{j}=\frac{1}{m}%
\mathop{\rm Im}%
\left[ f^{\ast }\hat{S}\left( \hat{S}\cdot \nabla f\right) \right] -\frac{1}{%
m}%
\mathop{\rm Im}%
\left[ f^{\ast }\left( \nabla \times \hat{S}\right) \times \hat{S}f\right] 
\]
where $\hat{S}$ is some unit vector. Angular momentum that results from this
current is zero, but if additional current is added that results from the
second term in (\ref{rospin}) one obtains 
\[
\vec{j}=\frac{1}{m}%
\mathop{\rm Im}%
\left[ f^{\ast }\hat{S}\left( \hat{S}\cdot \nabla f\right) \right] -\frac{1}{%
m}%
\mathop{\rm Im}%
\left[ f^{\ast }\left( \nabla \times \hat{S}\right) \times \hat{S}f\right] +%
\frac{1}{2m}\left( \nabla \times \vec{s}\right) f^{\ast }f 
\]
and by assuming that the last two cancel the resulting current produces
angular momentum. This current must obey the continuity equation 
\[
\partial _{t}\left( f^{\ast }f\right) =-\nabla \cdot \vec{j}=-\frac{1}{m}%
\mathop{\rm Im}%
\left[ f^{\ast }\left( \hat{S}\cdot \nabla \right) \left( \hat{S}\cdot
\nabla \right) f\right] 
\]
and when equation (\ref{schrfr}) is taken into account then it can only be
satisfied if 
\[
\left( \hat{S}\cdot \nabla \right) \left( \hat{S}\cdot \nabla \right)
f=\Delta f 
\]
For arbitrary $\hat{S}$ this identity is not possible, which means that from
the simple parameterisation (\ref{roun}) one cannot obtain equation with the
property that it exhibits the spin explicitly.

The phase space density (\ref{roun}) can be generalized as

\begin{eqnarray}
\rho (\vec{r},\vec{p}.t)\; &=&\;\frac{1}{\pi ^{3}}\int d^{3}q\,e^{2i\vec{p}%
\cdot \vec{q}}\left[ f^{\ast }(\vec{r}+\vec{q},t)f(\vec{r}-\vec{q}%
,t)+g^{\ast }(\vec{r}+\vec{q},t)g(\vec{r}-\vec{q},t)\right]  \label{ro2} \\
&\equiv &\frac{1}{\pi ^{3}}\int d^{3}q\,e^{2i\vec{p}\cdot \vec{q}}F^{+}(\vec{%
r}+\vec{q},t)F(\vec{r}-\vec{q},t)  \nonumber
\end{eqnarray}
which is also in accordance with the uncertainty principle but allows much
greater freedom to satisfy previous requirements. Important step in this
goal is to note that if one writes 
\begin{equation}
F(\vec{r})=f(\vec{r})\left| 
\begin{array}{c}
\cos \frac{\theta }{2}\;e^{-i\phi /2} \\ 
\sin \frac{\theta }{2}\;e^{i\phi /2}
\end{array}
\right| \equiv f(\vec{r})U(\theta ,\phi )  \label{spinpar}
\end{equation}
where $f(\vec{r})$ is a scalar, then in the product 
\[
U^{+}AU=A_{1,1}\cos ^{2}\frac{\theta }{2}+A_{2,2}\sin ^{2}\frac{\theta }{2}%
+\left( A_{1,2}e^{-i\phi }+A_{2,1}e^{i\phi }\right) \sin \frac{\theta }{2}%
\cos \frac{\theta }{2} 
\]
the matrix elements $A_{i,j}$ can be chosen so that every Cartesian
component of the unit radial vector $\hat{r}$ can be reproduced. It can be
easily verified that with the matrices 
\[
A=\sigma _{x}=\left| 
\begin{array}{cc}
0 & 1 \\ 
1 & 0
\end{array}
\right| \;\;\;;\;\;A=\sigma _{y}=\left| 
\begin{array}{cc}
0 & -i \\ 
i & 0
\end{array}
\right| \;\;\;\;;\;\;\;A=\sigma _{z}=\left| 
\begin{array}{cc}
1 & 0 \\ 
0 & -1
\end{array}
\right| 
\]
any unit vector can be written as $\hat{r}=\sigma _{x}\hat{x}+\sigma _{y}\;%
\hat{y}+\sigma _{z}\;\hat{z}=\vec{\sigma}$, where the product with the
matrices $U^{+}$ and $U$ is omitted but it is implied. The symbols $\sigma
_{x},\sigma _{y}\;$and $\sigma _{z}$ are used because they traditionally
represent Pauli spin matrices, but this is also known as the Cayley-Klein
parameterisation\cite{gold} in the classical mechanics of rigid body
dynamics which was developed much earlier that the concept of the spin was
introduced into physics. By using spin matrices it can be shown that any
vector, and in particular momentum, can be decomposed as 
\begin{equation}
\vec{p}=(\vec{\sigma}\cdot \vec{p})\vec{\sigma}-i\;\vec{\sigma}\times \vec{p}%
=\vec{\sigma}(\vec{\sigma}\cdot \vec{p})+i\;\vec{\sigma}\times \vec{p}\;
\label{pspin}
\end{equation}
where from now on the angles in the matrix $U$ give direction of the spin.
This is alternative decomposition of the momentum to that in (\ref{pdec1}),
which results in the probability current 
\begin{equation}
\vec{j}=\frac{1}{m}%
\mathop{\rm Im}%
\left[ F^{+}(\vec{r},t)\vec{\sigma}\;\vec{\sigma}\cdot \nabla F(\vec{r},t)%
\right] -\frac{1}{2m}\nabla \times F^{+}(\vec{r},t)\vec{\sigma}F(\vec{r},t)
\label{spincurr}
\end{equation}

It can be shown that the angular momentum resulting from the current (\ref
{spincurr}) is zero, but if the phase space density is modified as in (\ref
{rospin}), which in this case is\footnote{%
It should be pointed out that this addition to the phase space density in no
way affects consistency with the uncertainty principle.} 
\begin{eqnarray}
\rho (\vec{r},\vec{p}.t)\; &=&\frac{1}{\pi ^{3}}\int d^{3}q\,e^{2i\vec{p}%
\cdot \vec{q}}F^{+}(\vec{r}+\vec{q},t)F(\vec{r}-\vec{q},t)+  \label{ro2j} \\
&&\frac{s}{2\pi ^{3}}\left( \nabla _{r}\times \nabla _{p}\right) \cdot \int
d^{3}q\,e^{2i\vec{p}\cdot \vec{q}}F^{+}(\vec{r}+\vec{q},t)\vec{\sigma}F(\vec{%
r}-\vec{q},t)  \nonumber
\end{eqnarray}
where $s$ is for the moment arbitrary, then the current (\ref{spincurr}) is
modified as 
\begin{equation}
\vec{j}=\frac{1}{m}%
\mathop{\rm Im}%
\left[ F^{+}(\vec{r},t)\vec{\sigma}\;\vec{\sigma}\cdot \nabla F(\vec{r},t)%
\right] -\frac{1-s}{2m}\nabla \times F^{+}(\vec{r},t)\vec{\sigma}F(\vec{r},t)
\label{spinc}
\end{equation}
and the resulting angular momentum is not zero. The current (\ref{spinc})
satisfies the continuity equation because the divergence of the second term
is identically zero, while for the first term one can prove that $\left( 
\vec{\sigma}\cdot \nabla \right) \left( \vec{\sigma}\cdot \nabla \right)
=\Delta $ and $F$ satisfies the equation (\ref{schrfr}). Therefore essential
for the continuity equation is the first term in (\ref{spinc}) but from the
equation (\ref{schrfr}) it is not possible to derive it. This is possible if
the equation (\ref{schrfr}) is written in the equivalent form, as the set 
\begin{equation}
\partial _{t}F=-\vec{\sigma}\cdot \nabla G\;\;\;\;\;;\;\;\;\;\;G=\frac{1}{2im%
}\vec{\sigma}\cdot \nabla F  \label{schrspin}
\end{equation}
when it is simple to show that the first term in the current (\ref{spinc})
is derived. In this way the equations are derived that explicitly
incorporate the spin of the particle.

Still the parameter $s$ in (\ref{spinc}) is arbitrary, but its value is
determined from the requirement that the total angular momentum of the
particle should be conserved. From the time derivative 
\[
d_{t}\vec{L}=\frac{1}{\pi ^{3}}d_{t}\int d^{3}r\;d^{3}p\;\vec{r}\times \vec{p%
}\;\int d^{3}q\,e^{2i\vec{p}\cdot \vec{q}}F^{+}(\vec{r}+\vec{q},t)F(\vec{r}-%
\vec{q},t) 
\]
where the function $F$ satisfies the set of equations (\ref{schrspin}), and
after somewhat lengthy but straightforward manipulations, which involves the
set (\ref{schrspin}) and (\ref{pspin}), one obtains 
\[
d_{t}\vec{L}=-\frac{1}{2\pi ^{3}}d_{t}\int d^{3}rd^{3}p\int d^{3}q\,e^{2i%
\vec{p}\cdot \vec{q}}\;F^{+}(\vec{r}+\vec{q},t)\vec{\sigma}F(\vec{r}-\vec{q}%
,t) 
\]
which means that the total angular momentum can be defined either as 
\[
\vec{J}=\int d^{3}r\;d^{3}p\;\;\frac{1}{\pi ^{3}}\int d^{3}q\,e^{2i\vec{p}%
\cdot \vec{q}}F^{+}(\vec{r}+\vec{q},t)\left[ \vec{r}\times \vec{p}+\frac{1}{2%
}\vec{\sigma}\right] F(\vec{r}-\vec{q},t) 
\]
or 
\[
\vec{J}=\int d^{3}r\;d^{3}p\;\vec{r}\times \vec{p}\;\rho (\vec{r},\vec{p}.t) 
\]
where the phase space density is given by (\ref{ro2j}), but choosing $s=%
\frac{1}{2}$. This means that the spin of the particle is $1/2$. In the
first definition of the total angular momentum one integrates in all but
spatial variables, in which case 
\[
\vec{J}=%
\mathop{\rm Re}%
\left[ \int d^{3}r\;F^{+}(\vec{r},t)\left( \frac{1}{i}\vec{r}\times \nabla +%
\frac{1}{2}\vec{\sigma}\right) F(\vec{r},t)\right] 
\]
from where one defines operator for the total angular momentum as 
\[
\hat{J}=\frac{1}{i}\vec{r}\times \nabla +\frac{1}{2}\vec{\sigma} 
\]
This is recognized as the total angular momentum operator for the particle
with a spin in the standard quantum treatment.

It is important to make one observation. If one starts from the set of
equations (\ref{schrspin}) it is not simple to conclude that the spin of the
particle is 1/2, because the continuity equation that one derives from it is 
\[
\partial _{t}F^{+}(\vec{r},t)F(\vec{r},t)=-\frac{1}{m}\nabla \cdot 
\mathop{\rm Im}%
\left[ F^{+}(\vec{r},t)\vec{\sigma}\;\vec{\sigma}\cdot \nabla F(\vec{r},t)%
\right] 
\]
from where the only choice for the probability current is 
\[
\vec{j}=\frac{1}{m}%
\mathop{\rm Im}%
\left[ F^{+}(\vec{r},t)\vec{\sigma}\;\vec{\sigma}\cdot \nabla F(\vec{r},t)%
\right] 
\]
There is no way one can deduce the remainder in (\ref{spinc}), and in fact
one would conclude that the spin is 1, if it is calculated as the average 
\[
\vec{S}=m\int d^{3}r\;\vec{r}\times \vec{j} 
\]

\section{Charge with a spin in electromagnetic field}

In the previous section it was shown how to obtain equation for a free
particle which explicitly exhibits the spin. It is given by (\ref{schrspin}%
), but it is important to extend it for a charged particle in the
electromagnetic field. One starts from the Liouville equation for a particle
with a charge $e$ 
\begin{equation}
\partial _{t}\rho +\frac{1}{m}\vec{p}\cdot \nabla _{r}\rho +\left[ \vec{W}+%
\frac{e}{mc}\nabla (\vec{p}\vec{A})-\frac{e}{mc}(\vec{p}\nabla )\vec{A}%
\right] \cdot \nabla _{p}\rho =0  \label{liem}
\end{equation}
where $\Phi $ is scalar and $\vec{A}$ is vector potential, and $\vec{W}%
=-e\;\nabla \Phi -\frac{e}{c}\partial _{t}\vec{A}$. In order to select the
phase space density that is in accordance with the uncertainty principle,
and that resulting equations are for the spin $1/2$ particle, one starts
from the parametrization 
\[
\rho (\vec{r},\vec{p},t)=\frac{1}{\pi ^{3}}\int d^{3}q\;e^{2i\vec{q}\cdot (%
\vec{p}+\frac{e}{c}\vec{A})}F^{+}(\vec{r}+\vec{q},t)F(\vec{r}-\vec{q},t) 
\]
which is for a particle with no spin. It differs from (\ref{ro2}) in one
important aspect: the momentum is modified according to the recipe of
Section II (see in particular equation (\ref{roh}) ), in order to
accommodate its change when the charged particle is placed in the EM\ field.
If this parameterisation is placed into the Liouville equation (\ref{liem})
then by a set of transformations one finds the equation that $F$ satisfies.
They are quite tedious and lengthy, for example it can be shown that 
\begin{eqnarray*}
\pi ^{3}\;\vec{p}\cdot \nabla _{r}\rho &=&-\frac{e}{c}\int d^{3}q\;e^{2i\vec{%
q}\cdot (\vec{p}+\frac{e}{c}\vec{A})}\;\nabla _{r}\left[ \vec{q}\cdot \vec{A}%
\right] \cdot \nabla _{q}F^{+}(\vec{r}+\vec{q},t)F(\vec{r}-\vec{q},t)- \\
&&\frac{2ie^{2}}{c^{2}}\int d^{3}q\;e^{2i\vec{q}\cdot (\vec{p}+\frac{e}{c}%
\vec{A})}\vec{A}\cdot \nabla _{r}\left[ \vec{q}\cdot \vec{A}\right] F^{+}(%
\vec{r}+\vec{q},t)F(\vec{r}-\vec{q},t)- \\
&&\frac{1}{2i}\int d^{3}q\;e^{2i\vec{q}\cdot (\vec{p}+\frac{e}{c}\vec{A}%
)}\nabla _{q}\cdot \nabla _{r}F^{+}(\vec{r}+\vec{q},t)F(\vec{r}-\vec{q},t)-
\\
&&\frac{e}{c}\int d^{3}q\;e^{2i\vec{q}\cdot (\vec{p}+\frac{e}{c}\vec{A})}%
\vec{A}\cdot \nabla _{r}F^{+}(\vec{r}+\vec{q},t)F(\vec{r}-\vec{q},t)
\end{eqnarray*}
but one eventually ends with an expression for the Liouville equation that
can be considerably simplified if it is assumed that the scalar potential $%
\Phi $ is a quadratic function of coordinates and the vector potential $\vec{%
A}$ is linear. In such a case one can write 
\begin{equation}
\vec{q}\cdot \nabla \Phi (\vec{r})=\frac{1}{2}\Phi (\vec{r}+\vec{q})-\frac{1%
}{2}\Phi (\vec{r}-\vec{q})\;\;\;\;;\;\;\;\;\vec{A}(\vec{r})=\frac{1\;}{2}%
\vec{A}(\vec{r}+\vec{q})+\frac{1}{2}\vec{A}(\vec{r}-\vec{q})  \label{restr}
\end{equation}
and the Liouville equation is satisfied if $F$ satisfies 
\begin{equation}
i\;\partial _{t}F(\vec{r},t)+\frac{1}{2m}\left( \nabla -\frac{ie}{c}\bar{A}%
\right) ^{2}F(\vec{r},t)+e\Phi F(\vec{r},t)=0  \label{spin0}
\end{equation}

The equation for $F$ is for the charge with no spin, and to include the spin
one would follow the same procedure as in the previous section. However,
this simple extension is not straightforward and before doing that it is
necessary to discuss ordinary classical analysis of the time evolution of
the phase space density (\ref{rospin}) when EM force is included.

\subsection{Dynamics without uncertainty principle}

For a free particle with a spin the phase space density is given by (\ref
{rospin}), but when EM field is included then for $t=0$ its logical
extension would be 
\begin{equation}
\rho _{0}(\vec{r},\vec{p})\;=\rho _{in}(\vec{r},\vec{p}+\frac{e}{c}\vec{A})+%
\frac{\vec{s}}{2}\cdot \left( \nabla _{r}\times \nabla _{p}\right) \rho
_{in}(\vec{r},\vec{p}+\frac{e}{c}\vec{A})  \label{rosp0}
\end{equation}
and for arbitrary time 
\[
\rho (\vec{r},\vec{p},t)\;=\rho _{in}\left[ \vec{f},\vec{g}+\frac{e}{c}\vec{A%
}(\vec{f})\right] +\frac{\vec{s}}{2}\cdot \left( \nabla _{f}\times \nabla
_{g}\right) \rho _{in}\left[ \vec{f},\vec{g}+\frac{e}{c}\vec{A}(\vec{f})%
\right] 
\]
where the vectors $\vec{f}$ and $\vec{g}$ were defined in (\ref{trajh}), but
here they have functional dependence $\vec{f}(\vec{r},\vec{p},-t)$ and $\vec{%
g}(\vec{r},\vec{p},-t)$, respectively. The first term is not of immediate
interest, however the spin term, the second one, is. It will be analysed for
the simplest example, charge in a homogeneous magnetic field. The solution
for the classical trajectories was obtained in Section II, summarized in (%
\ref{trajh}), and therefore the spin can be calculated from 
\[
\vec{S}=\frac{1}{2}\int d^{3}r\;d^{3}p\;\vec{r}\times \vec{p}\;\vec{s}\cdot
\left( \nabla _{f}\times \nabla _{g}\right) \rho _{in}\left[ \vec{f},\vec{g}+%
\frac{e}{c}\vec{A}(\vec{f})\right] 
\]
To calculate the integral one needs to transform $\left( \nabla _{f}\times
\nabla _{g}\right) $ into the derivatives with respect to the coordinates
and components of the momentum, which is straightforward from the knowledge
of the vectors $\vec{f}$ and $\vec{g}$, but the result is somewhat
complicated. In short, one gets for the spin 
\[
\vec{S}=\left[ s_{x}+s_{xy}\cos (\alpha -\omega t)\right] \;\hat{x}+\left[
s_{y}+s_{xy}\sin (\alpha -\omega t)\right] \;\hat{y}+s_{z}\;\cos (\omega t)\;%
\hat{z} 
\]
where $s_{xy}$ is projection of the spin$\;\vec{s}$ in the x-y plane. Its
typical feature is that its modulus is not preserved, but precesses in the
x-y plane with the frequency $\omega =eh_{0}/(mc)$, and its z component also
oscillates. The solution appears acceptable, except that it does not conform
with the notion of the spin should be, e.g. it is expected that its modulus
should be preserved. Namely, if the spin should have any meaning then its
value must not depend on the circumstances as described. It can be shown on
another example that this definition of the spin suffers from further
serious drawbacks, e.g. if perpendicular to the magnetic field one applies
say a constant force then the time dependence of the spin, as it is defined,
would have a more complicated form. Yet the spin, if it is defined as a
property of the phase space density that depends on the magnetic field only,
should not be affected by such translation. Therefore a more elaborate time
dependence of the phase space density is required, which is more complicated
than the one based on propagating (\ref{rosp0}).

One possible generalization of the phase space density is 
\begin{equation}
\rho (\vec{r},\vec{p},t)=\rho _{0}(\vec{r},\vec{p},t)+\frac{1}{2}\left(
\nabla _{r}\times \nabla _{p}\right) \cdot \left[ \vec{s}(\vec{r},\vec{p}%
,t)\rho _{0}(\vec{r},\vec{p},t)\right]   \label{rospemcl}
\end{equation}
where $\vec{s}(\vec{r},\vec{p},t)$ will be called the spin function, and the
spin is now defined as the average 
\[
\vec{S}=\int d^{3}r\;d^{3}p\;\vec{s}(\vec{r},\vec{p},t)\rho _{0}(\vec{r},%
\vec{p},t)
\]
This parameterisation restricts considerably the possible phase space
densities, and its justification is possible a posterior, i.e. when the
uncertainty principle is introduced. Therefore in classical mechanics there
is no justification for such parameterisation, but it is entirely consistent
with the classical principles, as long as one finds the dynamics equations
that produce it. The parameterisation (\ref{rospemcl}) is possible at all
times if the equation for the spin function is found, and to find it simpler
problem will be considered: charge in the coordinate independent magnetic
field. In this case the spin function is only \ time dependent, and the
total time derivative of the phase space density is 
\[
d_{t}\rho (\vec{r},\vec{p},t)=\frac{1}{2}\partial _{t}\vec{s}(t)\cdot \left(
\nabla _{r}\times \nabla _{p}\right) \rho _{0}(\vec{r},\vec{p},t)+\frac{e}{%
2mc}\vec{s}(t)\cdot \left[ \nabla _{r}\times \left( \nabla _{p}\times \vec{H}%
\right) \right] \rho _{0}(\vec{r},\vec{p},t)
\]
where it was taken into account that $\rho _{0}(\vec{r},\vec{p},t)$
satisfies the Liouville equation 
\begin{equation}
\partial _{t}\rho _{0}+\frac{1}{m}\vec{p}\cdot \nabla _{r}\rho _{0}+\left[ 
\vec{W}+\frac{e}{mc}\vec{p}\times \vec{H}\right] \cdot \nabla _{p}\rho _{0}=0
\label{lisp0}
\end{equation}
and that $d_{t}\left( \nabla _{r}\times \nabla _{p}\right) =\nabla
_{r}\times \left( \nabla _{p}\times \vec{H}\right) +\left( \nabla _{r}\times
\nabla _{p}\right) d_{t}$ (the proof of this is not given, but can be
checked relatively easily). By slight rearrangement of the terms one gets
that if 
\begin{equation}
d_{t}\vec{s}(t)=\frac{e}{mc}\vec{s}(t)\times \vec{H}  \label{speq}
\end{equation}
then 
\[
d_{t}\rho (\vec{r},\vec{p},t)=-\frac{e}{2mc}\vec{s}(t)\cdot \left[ \nabla
_{p}\times \left( \vec{H}\times \nabla _{r}\right) \right] \rho _{0}(\vec{r},%
\vec{p},t)
\]
which violates the basic principle on which dynamics of the phase space
density is based on. i.e. the right side should be zero (Liouville
condition). This, however, appears a difficulty but in fact it is not, if
certain conditions are met. The basic one is that the total phase space
density is conserved, which means that 
\[
\int d^{3}r\;d^{3}p\;d_{t}\rho (\vec{r},\vec{p},t)=0
\]
This is indeed the case, but it is not the only condition. For example, from
the integral 
\[
\int d^{3}r\;d^{3}p\;\vec{p}\;d_{t}\rho (\vec{r},\vec{p},t)=-\frac{e}{2mc}%
\int d^{3}r\;d^{3}p\;\vec{p}\;\vec{s}(t)\cdot \left[ \nabla _{p}\times
\left( \vec{H}\times \nabla _{r}\right) \right] \rho _{0}(\vec{r},\vec{p},t)
\]
one would have to obtain the same equations of motion as the ones from which
time dependence of $\rho _{0}(\vec{r},\vec{p},t)$ is calculated. Detailed
analysis shows that this is indeed the case. Likewise, for the angular
momentum one would obtain 
\[
\int d^{3}r\;d^{3}p\;\vec{r}\times \vec{p}\;d_{t}\rho (\vec{r},\vec{p}%
,t)=\partial _{t}\vec{s}(t)-\frac{e}{2mc}\vec{s}(t)\times \vec{H}=\frac{e}{%
2mc}\vec{s}(t)\times \vec{H}
\]
from which the spin equation (\ref{speq}) is obtained, which shows self
consistency of the extended Liouville condition.

From the previous analysis one obtains that dynamics of the phase space
density is obtained from the set of two equations: one is (\ref{lisp0}) and
the other is (\ref{speq}). The problem, however, is how the extension is
made when the magnetic field is coordinate dependent? Analysis is more
complicated than the previous one but the idea is the same, and the final
result is the set of equations for dynamics of a particle with the spin 
\begin{equation}
d_{t}\vec{p}=\vec{W}+\frac{e}{mc}\vec{p}\times \vec{H}+\frac{e}{mc}\nabla
_{H}\left( \vec{s}\cdot \vec{H}\right) \;\;\;;\;\;d_{t}\vec{s}(t,\vec{r})=%
\frac{e}{mc}\vec{s}(t,\vec{r})\times \vec{H}\;  \label{cleqspin}
\end{equation}
from which the Liouville equation (\ref{lisp0}) modifies into 
\[
\partial _{t}\rho _{0}+\frac{1}{m}\vec{p}\cdot \nabla _{r}\rho _{0}+\left[ 
\vec{W}+\frac{e}{mc}\vec{p}\times \vec{H}+\frac{e}{mc}\nabla _{H}\left( \vec{%
s}\cdot \vec{H}\right) \right] \cdot \nabla _{p}\rho _{0}=0 
\]
where the index of the gradient indicates that spatial derivatives are taken
only for the magnetic field (the spin function is now coordinate dependent).

\subsection{Dynamics with uncertainty principle}

Previous analysis defined dynamics for a particle with a spin in EM field,
but important ingredient was omitted: it is not consistent with the
uncertainty principle. In order to make it the phase space density should be
parametrized as 
\begin{eqnarray}
\rho (\vec{r},\vec{p}.t)\; &=&\frac{1}{\pi ^{3}}\int d^{3}q\,e^{2i\vec{q}%
\cdot (\vec{p}+\frac{e}{c}\vec{A})}F^{+}(\vec{r}+\vec{q},t)F(\vec{r}-\vec{q}%
,t)+  \label{rospinem} \\
&&\frac{1}{4\pi ^{3}}\left( \nabla _{r}\times \nabla _{p}\right) \cdot \int
d^{3}q\,e^{2i\vec{q}\cdot (\vec{p}+\frac{e}{c}\vec{A})}F^{+}(\vec{r}+\vec{q}%
,t)\vec{\sigma}F(\vec{r}-\vec{q},t)  \nonumber
\end{eqnarray}
which is a straightforward generalization of (\ref{ro2j}). The function $F(%
\vec{r},t)$ can be parametrized as (\ref{spinpar}), but now having more
general functional character 
\begin{equation}
F(\vec{r},t)=f(\vec{r},t)U(\vec{r},t)  \label{fu}
\end{equation}
where $U^{+}(\vec{r},t)U(\vec{r},t)=I=unit\;matrix$. However, analysis
simplifies considerably if again it is assumed that magnetic field is
coordinate independent, which is also in accordance with the restriction (%
\ref{restr}). In this case one can write 
\begin{eqnarray*}
\rho (\vec{r},\vec{p}.t)\; &=&\frac{1}{\pi ^{3}}\int d^{3}q\,e^{2i\vec{q}%
\cdot (\vec{p}+\frac{e}{c}\vec{A})}f^{\ast }(\vec{r}+\vec{q},t)f(\vec{r}-%
\vec{q},t)+ \\
&&\frac{1}{2\pi ^{3}}\left( \nabla _{r}\times \nabla _{p}\right) \cdot \vec{s%
}(t)\int d^{3}q\,e^{2i\vec{q}\cdot (\vec{p}+\frac{e}{c}\vec{A})}f^{\ast }(%
\vec{r}+\vec{q},t)f(\vec{r}-\vec{q},t)
\end{eqnarray*}
which is in the form (\ref{rospemcl}). The equation for the spin function is
(\ref{speq}), which implies that 
\[
d_{t}U^{+}(t)\;\vec{\sigma}U(t)=\frac{e}{mc}U^{+}(t)\vec{\sigma}\times \vec{H%
}U(t) 
\]
From (\ref{pspin}) one can write 
\[
\;\vec{\sigma}\times \vec{H}=\frac{1}{2i}\left[ (\vec{\sigma}\cdot \vec{H})%
\vec{\sigma}-\vec{\sigma}(\vec{\sigma}\cdot \vec{H})\right] 
\]
in which case the equation for the function $U(t)$ is 
\[
i\;d_{t}U(t)=-\frac{e}{2mc}\vec{\sigma}\cdot \vec{H}\;U(t) 
\]
while that for the function $f(\vec{r},t)$ is (\ref{spin0}). The product of
the two functions satisfies the equation 
\begin{equation}
i\;\partial _{t}F=-\frac{1}{2m}\left( \nabla -\frac{ie}{c}\bar{A}\right)
^{2}F-\frac{e}{2cm}\vec{H}\cdot \vec{\sigma}F+e\Phi F(\vec{r},t)
\label{spineq}
\end{equation}
It can be easily shown that this equation is derived from the set

\begin{equation}
\partial _{t}F=-\vec{\sigma}\cdot \left( \nabla -\frac{ie}{c}\bar{A}\right)
G+e\Phi F(\vec{r},t)\;\;\;\;;\;\;\;\;\;G=\frac{1}{2im}\vec{\sigma}\cdot
\left( \nabla -\frac{ie}{c}\bar{A}\right) F\;  \label{fgspina}
\end{equation}
which is straightforward generalization of the set (\ref{schrspin}).

Generalization of the equation (\ref{spineq}) for the magnetic field that is
not homogeneous is straightforward, but connection with the classical set of
equations is lost, because of the condition (\ref{restr}) which is not
satisfied. Therefore, the connection is only approximate, and to what degree
will be checked on one example.

\section{Gauge invariance and two examples}

One very important aspect of the previous analysis was entirely neglected,
and it is imperative to mention it. The gauge invariance was not checked,
which is a very important condition that should be satisfied. In the
standard quantum treatment the problem appears to have been solved, but
simple example indicates the contrary. For example if one would want to
obtain momentum distribution for a charge in, say, static magnetic field
then according to standard definition 
\[
g(\vec{p})=\int d^{3}r\;f(\vec{r})\;e^{-i\vec{r}\cdot \vec{p}} 
\]
which is not gauge invariant. One could try various amendments to obtain the
invariance but non of them are satisfying. In this work gauge invariance
comes naturally, and the problem of the momentum distribution is relatively
easily solved. One should go back to the observation in Section II that
momentum in the EM field is different from its value before the field is
turned on. The relationship is summarized in (\ref{initvh}) (for simplicity
static EM field is considered), which means that because momentum $\vec{p}$
is gauge invariant then momentum $\vec{P}$ is not. It follows from the
relationship (\ref{initvh}) that in order to satisfy this requirement then
the gauge $\vec{A}\rightarrow \vec{A}+\nabla \Theta $ implies $\vec{P}%
\rightarrow \vec{P}-\nabla \Theta $. This, however, implies that the
parametrization (\ref{rosp0}) is gauge invariant (small $\vec{p}$ is used
for convenience, but it is in fact capital $\vec{P}$), and that under the
gauge transformation the equation (\ref{spineq}) transforms as 
\[
i\;\partial _{t}F=-\frac{1}{2m}\left( \nabla +\frac{ie}{c}\nabla \Theta -%
\frac{ie}{c}\bar{A}-\frac{ie}{c}\nabla \Theta \right) ^{2}F-\frac{e}{2cm}%
\vec{H}\cdot \vec{\sigma}F+e\Phi F(\vec{r},t) 
\]
and it is invariant. However, instead of transforming momentum one can
formally assume that the wave function transforms as $F\rightarrow Fe^{\frac{%
ie}{c}\Theta }$, in which case gauge invariance is satisfied without
assuming transformation properties of $\nabla $.

One example in which momentum distribution will be calculated is
generalization of the problem that was treated in Section II. In addition to
charge being in homogeneous magnetic field it will be assumed that it is
also in the field of three dimensional harmonic oscillator, and before
switching on magnetic field the charge is assumed to be in the ground state
of that oscillator. The problem will be solved by classical equations of
motion, and from the previous analysis it follows that the result is
identical to that obtained by solving equation (\ref{spineq}). In this
example dynamics of the spin is not analyzed, only the probability densities 
\[
P(\vec{r},t)=\int d^{3}p\;\rho (\vec{r},\vec{p},t)\;\;\;;\;\;Q(\vec{p}%
,t)=\int d^{3}r\;\rho (\vec{r},\vec{p},t)\; 
\]
The equations from which dynamics of the phase space density is deduced are 
\[
d_{t}\vec{p}=-\nabla \Phi +\frac{e}{mc}\vec{p}\times \vec{H}+\frac{e}{mc}%
\nabla _{H}\left( \vec{s}\cdot \vec{H}\right) \;\;\;;\;\;\;d_{t}\vec{s}(t)=%
\frac{e}{mc}\vec{s}(t)\times \vec{H} 
\]
and for homogeneous magnetic field the last term in the first equation is
zero. The potential for harmonic oscillator is. $\Phi =\frac{1}{2}m\omega
_{0}r^{2}$, and if one does re-scaling $\vec{s}/\hbar \rightarrow \vec{s}$, $%
ct\varkappa \rightarrow t$ and $\vec{r}\varkappa \rightarrow \vec{r}$, where 
$\varkappa =mc/\hbar $, then the first equation is (the spin equation is not
analyzed) 
\[
d_{t}^{2}\vec{r}=\omega \;d_{t}\vec{r}\times \hat{z}-\omega _{0}^{2}\;\vec{r}
\]
where now oscillator frequency $\omega _{0}$ stands for $\omega _{0}\hbar
/(mc^{2})$ and the cyclotron frequency $\omega $ for $eH_{0}\hbar
/(m^{2}c^{3})$. For simplicity this equation is solved in the x-y plane
(magnetic field does not have effect on the motion along the z direction),
and the solution is 
\begin{eqnarray*}
\vec{r} &=&\vec{r}_{0}\left[ \cos \left( \frac{t\omega }{2}\right) \cos
\left( \frac{t}{2}\sqrt{\omega ^{2}+4\omega _{0}^{2}}\right) +\frac{\omega }{%
\sqrt{\omega ^{2}+4\omega _{0}^{2}}}\sin \left( \frac{t\omega }{2}\right)
\sin \left( \frac{t}{2}\sqrt{\omega ^{2}+4\omega _{0}^{2}}\right) \right] +
\\
&&\hat{z}\times \vec{r}_{0}\left[ \sin \left( \frac{t\omega }{2}\right) \cos
\left( \frac{t}{2}\sqrt{\omega ^{2}+4\omega _{0}^{2}}\right) -\frac{\omega }{%
\sqrt{\omega ^{2}+4\omega _{0}^{2}}}\cos \left( \frac{t\omega }{2}\right)
\sin \left( \frac{t}{2}\sqrt{\omega ^{2}+4\omega _{0}^{2}}\right) \right] +
\\
&&\frac{2\vec{v}_{0}}{\sqrt{\omega ^{2}+4\omega _{0}^{2}}}\cos \left( \frac{%
t\omega }{2}\right) \sin \left( \frac{t}{2}\sqrt{\omega ^{2}+4\omega _{0}^{2}%
}\right) +\frac{2\hat{z}\times \vec{v}_{0}}{\sqrt{\omega ^{2}+4\omega
_{0}^{2}}}\sin \left( \frac{t\omega }{2}\right) \sin \left( \frac{t}{2}\sqrt{%
\omega ^{2}+4\omega _{0}^{2}}\right)
\end{eqnarray*}
where $\vec{r}_{0}$ and $\vec{v}_{0}$ are initial position and velocity of
the charge, in the x-y plane. According to prescription (\ref{roh}) one
obtains the phase space density at any time, and result is quite
complicated. However, probability densities $P(\vec{r},t)$ and$\;Q(\vec{p}%
,t) $ are simpler, and given by 
\[
P(\vec{r},t)=\frac{2}{\pi d^{2}}\frac{4+q^{2}}{\Delta _{r}}\exp \left[ -%
\frac{2(4+q^{2})}{d^{2}\Delta _{r}}r^{2}\right] \;\;;\;\;Q(\vec{p},t)=\frac{%
4d^{2}}{\pi }\frac{4+q^{2}}{\Delta _{p}}\exp \left[ -\frac{4d^{2}(4+q^{2})}{%
\Delta _{p}}p^{2}\right] 
\]
where 
\begin{eqnarray*}
\Delta _{r} &=&8+q^{2}+q^{2}\cos \left( t\sqrt{\frac{4+q^{2}}{d^{4}}}\right)
\\
\Delta _{p} &=&16+10q^{2}+q^{4}-2q^{2}\cos \left( t\sqrt{\frac{4+q^{2}}{d^{4}%
}}\right)
\end{eqnarray*}
where $q=\omega /\omega _{0}$ and $d=1/\sqrt{\omega _{0}}$ is the width of
the ground state of the harmonic oscillator. Interesting limit is $q\gg 1$,
in which case the width of 
\[
P(\vec{r},t)\sim \frac{1}{1+\cos \left( t\omega \right) }\exp \left[ -\frac{%
2r^{2}}{d^{2}\left[ 1+\cos \left( t\omega \right) \right] }\right] 
\]
oscillates between $d$ (unperturbed oscillator) and (nearly) zero. On the
other hand, momentum distribution has the limiting form 
\[
Q(\vec{p},t)\sim \frac{1}{q^{2}}\exp \left[ -\frac{4d^{2}}{q^{2}}p^{2}\right]
\]
which is (nearly) time independent, but it is very wide.

In the second example dynamics in the inhomogeneous magnetic field of the
charge with a spin will be analyzed. For the vector potential it is assumed 
\[
\vec{A}=h_{1}z\left( -y\;\hat{x}+x\;\hat{y}\right) 
\]
and the magnetic field is 
\[
\vec{H}=-h_{1}x\;\hat{x}-h_{1}y\;\hat{y}+2h_{1}z\;\hat{z} 
\]
In the scaling defined earlier the classical set of equations is 
\begin{equation}
\stackrel{..}{\vec{r}}=\varepsilon \left[ \stackrel{.}{\vec{r}}\times \vec{H}%
+\frac{1}{2}\nabla _{H}\left( \vec{s}\cdot \vec{H}\right) \right]
\;\;\;;\;\;\;\partial _{t}\vec{s}(t,\vec{r})=\varepsilon \;\vec{s}(t,\vec{r}%
)\times \vec{H}  \label{cl}
\end{equation}
while the equation (\ref{spineq}) (quantum equation) is 
\begin{equation}
i\;\partial _{t}F=-\frac{1}{2}\left( \nabla -i\varepsilon \bar{A}\right)
^{2}F-\frac{\varepsilon }{2}\vec{H}\cdot \vec{\sigma}F  \label{qv}
\end{equation}
where $\varepsilon =e\hbar ^{2}h_{1}/(m^{3}c^{4})$, and in vector potential
the coefficient $h_{1}$ is omitted. The equations (\ref{cl}) and (\ref{qv}),
together with the appropriate initial conditions, determine dynamics of the
charge. As it was shown, if the magnetic field is coordinate independent,
but with arbitrary time dependence, the two sets produce identical results.
However, it is not clear that the two dynamics produce the same result if
this condition is not satisfied, and the purpose of the following analysis
so to check if this were the case.

In classical dynamics (which is based on the set of equations (\ref{cl}) )
one calculates time dependence of the phase space density (\ref{rospinem}),
and from parametrization (\ref{fu}) one assumes its general form 
\[
\rho (\vec{r},\vec{p},t)=\rho _{0}(\vec{r},\vec{p},t)+\frac{1}{2}\left(
\nabla _{r}\times \nabla _{p}\right) \cdot \vec{s}(t,\vec{r})\rho _{0}(\vec{r%
},\vec{p},t) 
\]
where 
\[
\rho _{0}(\vec{r},\vec{p},t)=\frac{1}{\pi ^{3}}\int d^{3}q\,e^{2i\vec{q}%
\cdot (\vec{p}+\varepsilon \vec{A})}f^{\ast }(\vec{r}+\vec{q},t)f(\vec{r}-%
\vec{q},t) 
\]
The dynamics of the particle is therefore determined if initial $\rho _{0}(%
\vec{r},\vec{p})$ is specified together with the orientation of the spin $%
\vec{s}(t=0,\vec{r})=\vec{s}_{0}$ (its modulus is fixed). From the phase
space density various quantities can be calculated, but one in particular is
the spin 
\begin{equation}
\vec{s}(t)=\int d^{3}r\;d^{3}p\;\vec{r}\times \vec{p}\;\rho _{sp}(\vec{r},%
\vec{p},t)=\int d^{3}r\;d^{3}p\;\vec{s}(t,\vec{r})\;\rho _{0}(\vec{r},\vec{p}%
,t)  \label{spincl}
\end{equation}
In practical implementation one generates random initial $\vec{r}$ and $\vec{%
p}$ from the distribution $\left| \rho _{0}(\vec{r},\vec{p})\right| $ and
solves equations of motion (\ref{cl}). For $N$ randomly chosen initial
conditions the spin, for example, is approximately (for $N\rightarrow \infty 
$ the result is exactly the same as (\ref{spincl}) ) 
\[
\vec{s}(t)\approx \frac{\sum_{n=1}^{N}\vec{s}(t,\vec{r}_{n})\;%
\mathop{\rm Si}%
gn\left[ \rho _{0}(\vec{r}_{n},\vec{p}_{n})\right] }{\sum_{n=1}^{N}%
\mathop{\rm Si}%
gn\left[ \rho _{0}(\vec{r}_{n},\vec{p}_{n})\right] } 
\]
where $%
\mathop{\rm Si}%
gn\left[ \rho _{0}(\vec{r}_{n},\vec{p}_{n})\right] $ is the sign of the
initial phase space density for the $n-th$ pair of initial conditions.

Solution based on the equation (\ref{qv}) (quantum solution) is more
difficult to obtain. First, because of the cylindrical symmetry the equation
is transformed into the cylindrical coordinates, and if $F$ is parametrized
as 
\[
F_{n}=\frac{1}{\sqrt{r}}e^{in\phi }\left| 
\begin{array}{cc}
1 & 0 \\ 
0 & e^{i\phi }
\end{array}
\right| G_{n}(r,z,t)\;\;\;\;;\;\;\;n=0,\pm 1,\pm 2,.... 
\]
where $r$ is the radial distance in the x-y plane and $\phi $ is the
azimuthal angle, then the equation for $G_{n}(r,z)$ is angle independent. If
the initial $F$ is assumed to have the form 
\[
F=f_{0}(r,z)\left| 
\begin{array}{c}
\cos \frac{\alpha }{2}\;e^{-i\beta /2} \\ 
\sin \frac{\alpha }{2}\;e^{i\beta /2}
\end{array}
\right| 
\]
where $\alpha $ and $\beta $ are directional angles of the spin, polar and
azimuthal angles, respectively, then it is obvious that single $F_{n}$ is
not sufficient to describe dynamics. One needs $F_{0}$ and $F_{-1}$ to
describe it, and the initial values for $G_{n}(r,z,t)$ are 
\[
G_{0}(r,z,0)=\sqrt{r}f_{0}(r,z)\left| 
\begin{array}{c}
\cos \frac{\alpha }{2}\;e^{-i\beta /2} \\ 
0
\end{array}
\right| \;\;\;;\;\;\;G_{-1}(r,z,0)=\sqrt{r}f_{0}(r,z)\left| 
\begin{array}{c}
0 \\ 
\sin \frac{\alpha }{2}\;e^{i\beta /2}
\end{array}
\right| 
\]
and the overall solution is $F=F_{0}(r,z,t)+F_{-1}(r,z,t)$. In order to
simplify analysis it will be assumed that the spin is initially oriented
along the z direction, in which case only one equation needs to be solved,
for the index $n=0$. The equation for $G_{0}(r,z,t)$ is 
\[
i\;\partial _{t}G_{0}=-\frac{1}{2}\left[ \partial _{r}^{2}+\partial _{z}^{2}+%
\frac{1}{2r^{2}}\left( \sigma _{z}-\frac{1}{2}\right) -\varepsilon
^{2}z^{2}r^{2}\right] G_{0}+\frac{\varepsilon }{2}r\sigma _{x}G_{0}-\frac{1}{%
2}\varepsilon z\left( \sigma _{z}+I\right) G_{0} 
\]
and it was solved numerically by replacing partial spatial derivatives with
their finite difference approximate forms (care should be taken in
implementing this procedure for the radial coordinate, but the details are
omitted). In this case the equation is replaced by a matrix equation 
\begin{equation}
i\;d_{t}G_{0}(r_{n},z_{m},t)=\sum_{j,k}O\left(
r_{n},z_{m};r_{j},z_{k}\right) G_{0}(r_{j},z_{k},t)  \label{teq}
\end{equation}
which is solved numerically as a set of differential equations. The accuracy
of the solution depends on two factors. One is the step that is used for
approximating the derivatives, and the other is how well the boundary
conditions are satisfied. The limitation on the former is the amount of
computer time one has at disposal for calculation. Limitations from the
latter are in the bounds that are set on the range of coordinates, because
one chooses $z_{\min }$ and $z_{\max }$ for $z$, and $r_{\max }$ for $r$,
within which the solution is confined at all times. Implication is that at
these ends $G_{0}(r_{\max },z_{\min ,\max },t)=0$, which means that the
solution is accurate for all times when this condition is satisfied.

Explicit example was calculated for 
\[
f_{0}(r)=\frac{1}{\sqrt{d^{3}\pi ^{3/2}}}e^{-\frac{r^{2}}{2d^{2}}} 
\]
and the initial phase space density is 
\[
\rho _{0}(\vec{r},\vec{p})\;=\frac{1}{\pi ^{3}}e^{-\frac{r^{2}}{d^{2}}-d^{2}(%
\vec{p}+\varepsilon \vec{A})^{2}} 
\]
Initial orientation of the spin was assumed along the z axes ($\alpha =0$).
Numerical values for the parameters are $d=10$ and $\varepsilon =0.01$,
which were chosen so that the important effects are easily noticed, however
they are not necessarily easily realized in experiment. Thus for the
electron $d\approx 2.4\ast 10^{-11}m$ and $h_{1}\approx 10^{14}T/m$. This
field gradient is enormous and beyond means of any practical set up, but the
effects of it on dynamics of the charge with the spin are clearly manifested.

Two quantities were calculated from classical and quantum equations, one is
the spin, which in classical dynamics is defined as (\ref{spincl}) and in
quantum as 
\begin{equation}
\vec{s}(t)=\frac{1}{2}\int d^{3}r\;G_{0}^{+}\vec{\sigma}G_{0}  \label{spinqv}
\end{equation}
and the other is the probability density for finding the charge along the z
coordinate, and it is defined as 
\[
P(z,t)=\int d^{3}p\;dx\;dy\;\rho _{0}(\vec{r},\vec{p},t)=\int
dx\;dy\;G_{0}^{+}G_{0} 
\]
In the classical calculation $10^{5}$ trajectories were calculated to obtain
reasonable accurate phase space density. In the quantum calculations the
following parameters were chosen: $z_{\min }=-100$ and $z_{\max }=200$ with
the number of divisions of this interval $n_{z}=600$, and $r_{\max }=200$
with the number of divisions $n_{r}=200$ (total of 240000 equations in the
set (\ref{teq}) ). Large value of these parameters to obtain (reasonable)
accurate quantum results (as the rule for longer time evolution one needs
considerably larger set of equations (\ref{teq}) ) is an indication that
dynamics is dominated by some irregular process, about which it will be
learned shortly. Time dependence of the spin, which is always pointing in
the z direction, is shown in Figure 1, where the solid line shows quantum
result and broken line classical. The agreement is nearly perfect up to the
time $t=30\;(\approx 3.9\;10^{-20}\sec )$ when deviation sets in. The
possible origin of this deviation will be discussed shortly, but the result
in itself is interesting to emphasize. Both the classical and quantum
dynamics indicate (in more or less identical manner) that the spin does not
have a fixed value, but its modulus changes with interaction. In fact it
changes with the gradient of the field, because as soon as the field becomes
uniform the spin restores its original value $1/2$. It is interesting,
however, to note that the classical equation of motion for the spin function
in (\ref{cl}) conserves its modulus, and the same is true for the quantum
equation, but the spin, being either classical (\ref{spincl}) or quantum (%
\ref{spinqv}) average over the spin function, it is not conserved.

The probability density $P(z,t)$ for the time interval till $t=20$ is shown
in Figure 2, where the solid line is quantum result and broken is classical.
The difference between the two results is negligible. However, for the time
beyond $t=30$ it becomes noticeable, as shown in Figure 3. The essential
feature of the probability density is that it consists of two parts, the
central and stable peak around $z=0$ and its extension in the positive z
direction. The central peak is dominant, and the two calculations give
nearly identical results, however, the extension in quantum calculations is
(randomly) oscillatory while classical result appears as the average of it.
Some of the oscillatory structure in quantum result is manifestation of
numerically not converged calculation, but overall structure is reliable.
From the analysis of the previous section it is expected that classical and
quantum results are not identical, but the surprising finding is that
quantum is so difficult to obtain and it has such a (nearly) random
structure. Explanation for the discrepancy in Figure 1 can now be given. The
spin is the average over the spin function, whether classical or quantum,
and because quantum probability shows the oscillatory structure and
classical do not the averages cannot be expected to be the same. In fact the
deviation between the classical and quantum calculation of the spin starts
to be noticeable at the time when the quantum oscillations start to dominate
the probability density. In any case the exact agreement is not expected, as
derivations of the previous section indicates, and when this is taken into
account discrepancy between classical and quantum results is not fatal, in
fact it is quite good. So much so that it can be claimed that the spin has
classical and non relativistic origin, although precise dynamics of the
particle with a spin should be described by the set of equations (\ref
{fgspina}). However, classical description of the spin has one great
advantage, it offers understanding of the underlying processes that
contribute to the final results, shown in Figures 1, 2 and 3. They are
result of averaging over large number of trajectories, each with different
initial conditions for the coordinates and momentum (the spin is always
fixed). Few typical ones are shown in Figure 4, where time dependence of the
z coordinate of the particle, the z-th component of the spin function and
projection of the spin function in the x-y plane are shown. The modulus of
the spin function is always constant for each trajectory, but its x-y
projection is not zero. On the other hand, both quantum and classical
calculations indicate that the spin has only the z-th component, which means
that the zero x-y component of the spin is result of averaging. In fact in
classical calculations with a finite number of trajectories the x-y
component of the spin is not zero, but as its number is increased it
converges to zero. Another important result is shown in Figure 4. Each
trajectory is different, with no relationship between them. Even two very
close with the initial conditions soon become unrelated, and indication of
the underlying chaos in dynamics. This means that the ''chaotic''
oscillatory structure in Figure 3 is manifestation of the chaotic underlying
dynamics, and explanation why there was difficulty to obtain reasonable
accurate results.

\section{Summary}

Analysis of the spin, and calculation of examples, indicate that the spin is
classical in origin, although in summary it could be labeled as the
''fundamental parameter'' of the particle, rather than derivable from some
classical model like a spinning sphere. The reason why the spin should be
regarded as such is in the way how the spin function was introduced. By
parametrizing the phase space density as (\ref{rospemcl}), and enforcing
that it retains this a form in the course of time, has as the result that
the spin function is determined. It is a solution of an additional equation
for the dynamics of the particle, so that a complete classical set of
equations is (\ref{cleqspin}). The first equation is, without the spin
function term, the usual classical equation of motion, and by including this
term it is coupled to the second equation for the spin function. There is a
very profound impact of this dynamics on the phase space density. The spin
function is an additional degree of freedom for the classical particle,
which means that at a given point in time two, or more, classical
trajectories can cross (this is not possible for a spinless particle) in the
phase space. This is possible because trajectories are not determined by
only the position and momentum of the particle, but also by their spin
function. Therefore if for a given time $t$ two trajectories cross in the
phase space this means that their spin functions are different. As the
result the requirement that the value of the phase space density is
preserved in time, i.e. $d_{t}\rho (\vec{r},\vec{p},t)=0$, from which the
Liouville equation for the spinless particle is derived, is no longer true.
For the particle with the spin the Liouville equation is derived from the
inhomogeneous relationship $d_{t}\rho (\vec{r},\vec{p},t)=f(\vec{r},\vec{p}%
,t)$, but in order to preserve the physicality of the problem the function $%
f(\vec{r},\vec{p},t)$ must satisfy certain requirements. Those were
discussed in the section following the equation (\ref{rospemcl}). In this
respect the spin appears as a fundamental property of the particle, in
addition to its charge and mass, even in classical mechanics. However, the
observed value of the spin is the average over the spin function, and from
this point it is a quantity whose magnitude greatly depends on the forces
that act on the particle.

There is another point that needs a brief discussion. It is customary to
think not only that the spin does not have classical description, but also
that it is inherently a relativistic phenomenon, i.e. result of making
quantum theory relativistic, but in a specific manner as Dirac did. The
original intention of Dirac was not to look for the equation that describes
the particle with the spin. The intention was to make the Klein-Gordan
equation first order in the space-time derivatives, for the reasons that can
be found explained in any textbook on relativistic quantum theory. In doing
so four component wave function was introduced, and the equations were found
for their components. In the non relativistic limit these equations reduce
to (\ref{schrspin}) for a free particle and to (\ref{fgspina}) for a
particle that is in the EM field. However, in this work the pursuit was to
find equations for the particle with the spin, and it was explicitly shown
that they can be obtained if and only if one makes generalization (\ref{ro2}%
) of the phase space density. This is equivalent to requiring that the wave
function is not single but the four component object that produces the set (%
\ref{schrspin}), which is also first order in space-time derivatives, the
same as the Dirac equation. In conclusion, therefore, one can say that Dirac
derivation was accidental discovery of the spin, and because no alternative
arguments (Pauli spin equation is an ad-hoc derivation, and can only be
traced to the non relativistic limit of the Dirac equation) for deriving the
Dirac equation were found the only realistic conclusion was that the spin is
of the relativistic origin, with no true explanation in non relativistic
quantum theory (not to mention classical theory).

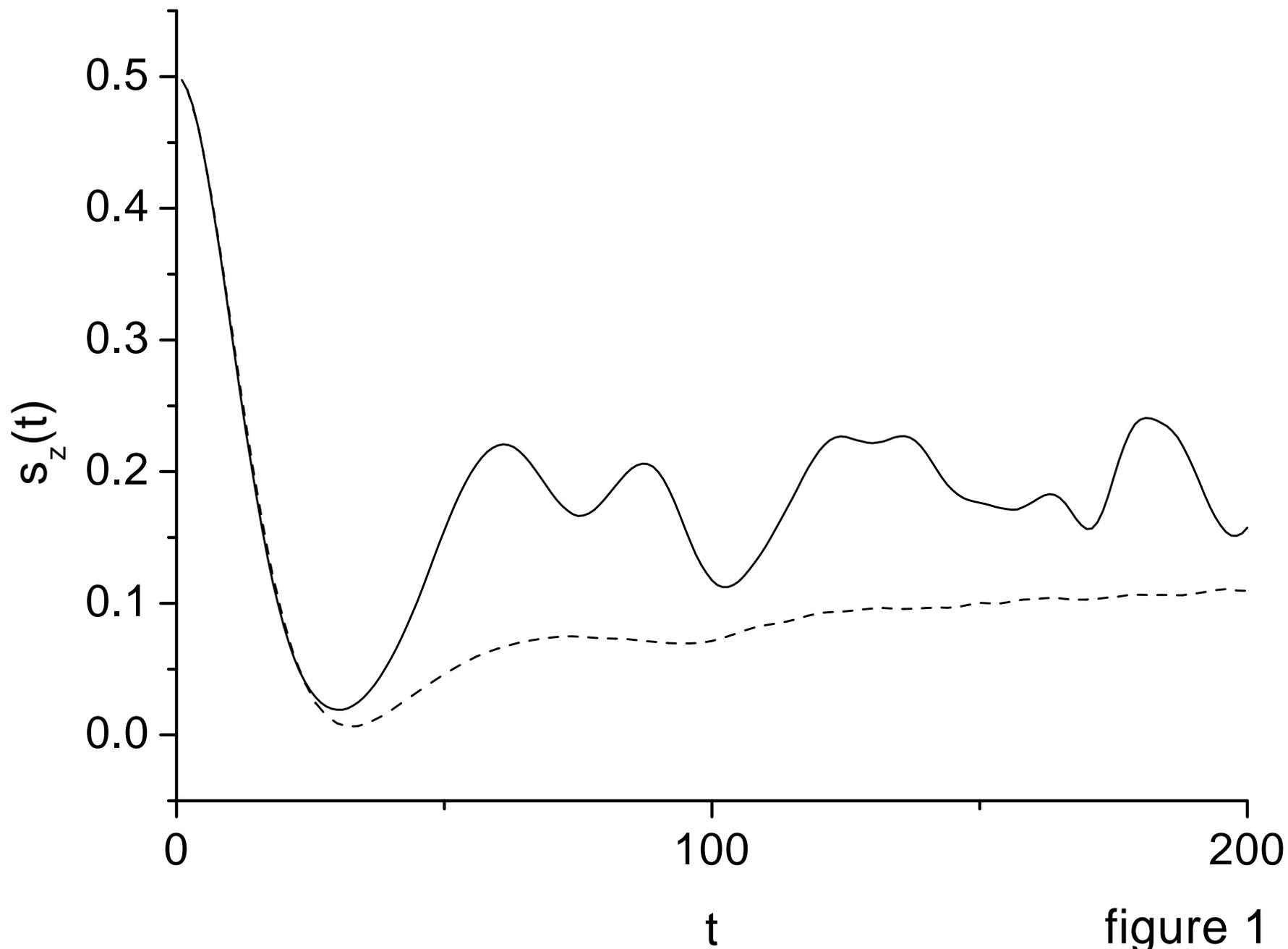
\begin{figure}[tbp]
\caption{Quantum (solid line) and classical (broken line) time dependence of
the spin of the spin 1/2 particle in inhomogeneous magnetic field, whose
axes of symmetry is z axes. Initially the spin of the particle is parallel
to the z axes, and in the course of time it remains so.}
\label{fig1}
\end{figure}

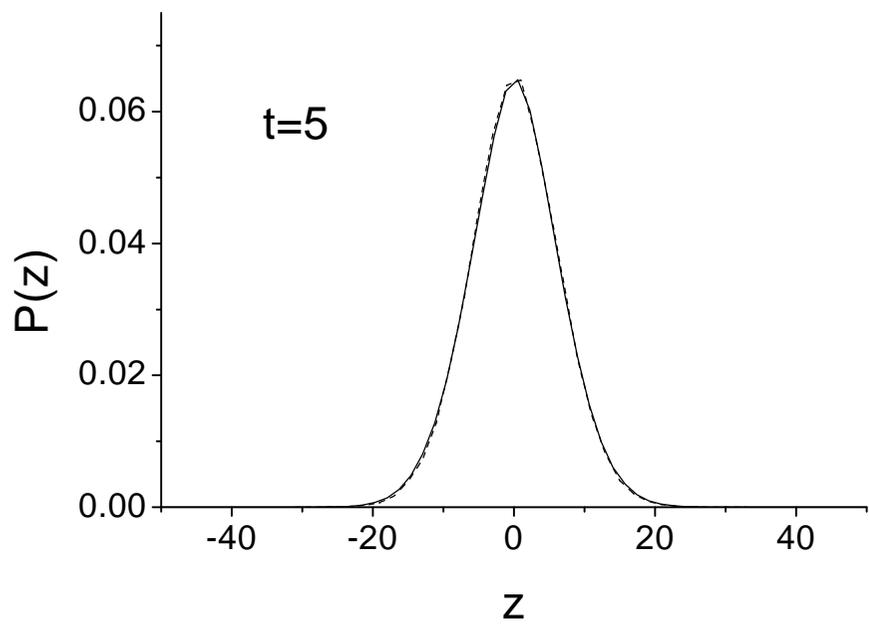
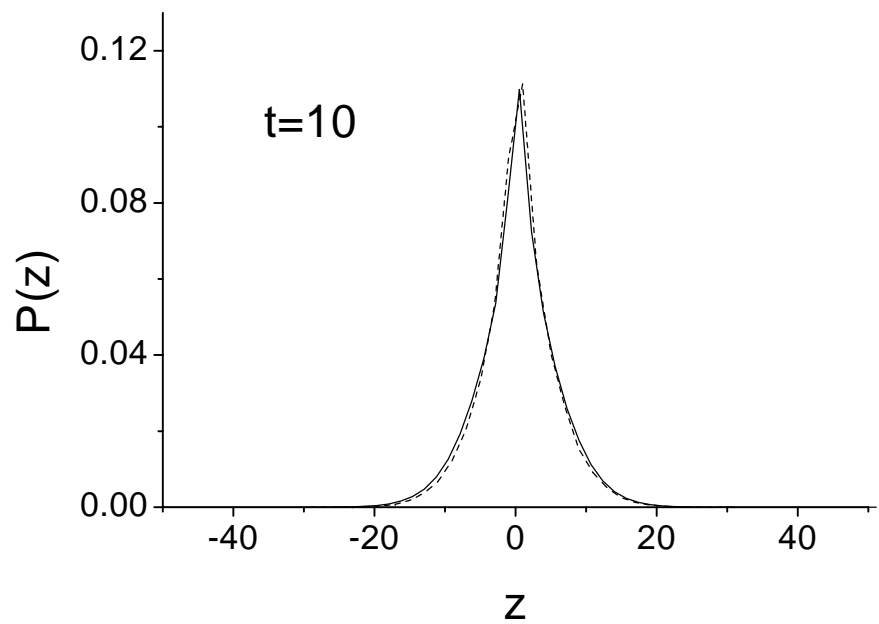
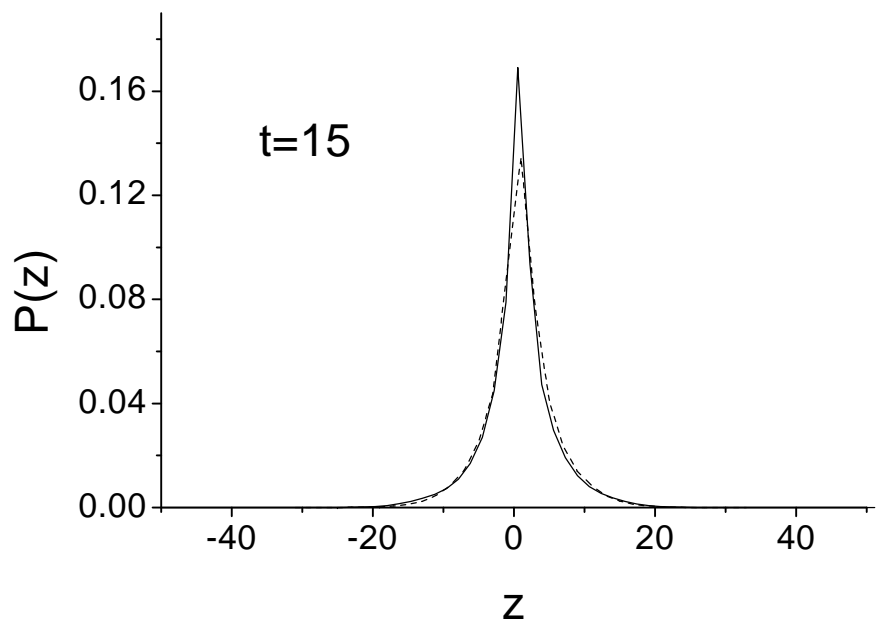
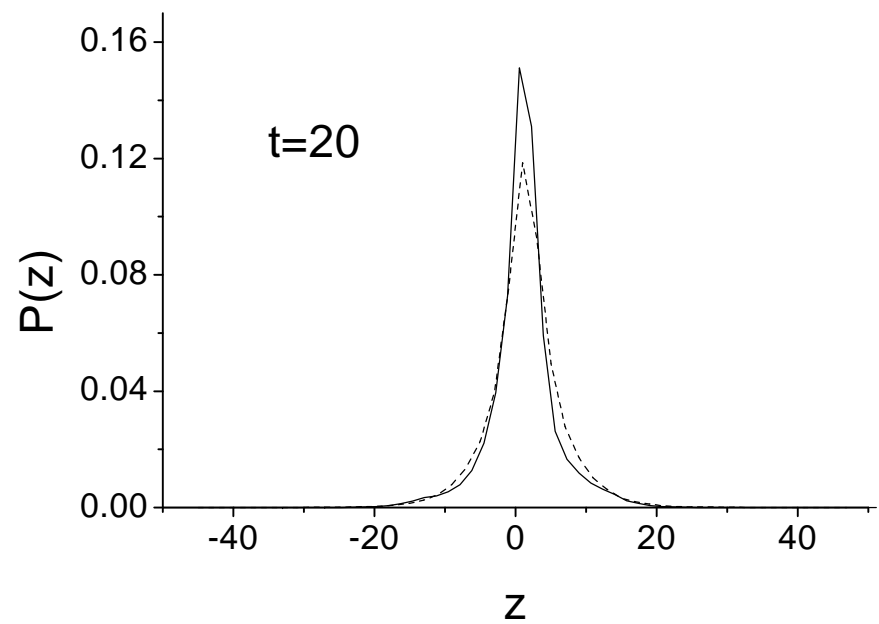
\begin{figure}[tbp]
\caption{Quantum (solid line) and classical (broken line) time evolution for
the probability density P(z,t), up to the time when quantum and classical
calculations for the spin of the particle start to separate, as shown in
Figure 1.}
\label{fig2}
\end{figure}

\begin{figure}[tbp]
\caption{ Quantum (solid line) and classical (broken line) time evolution
for the probability density P(z,t), beyond the time when quantum and
classical calculations for the spin of the particle produce nearly identical
results.}
\label{fig3}
\end{figure}

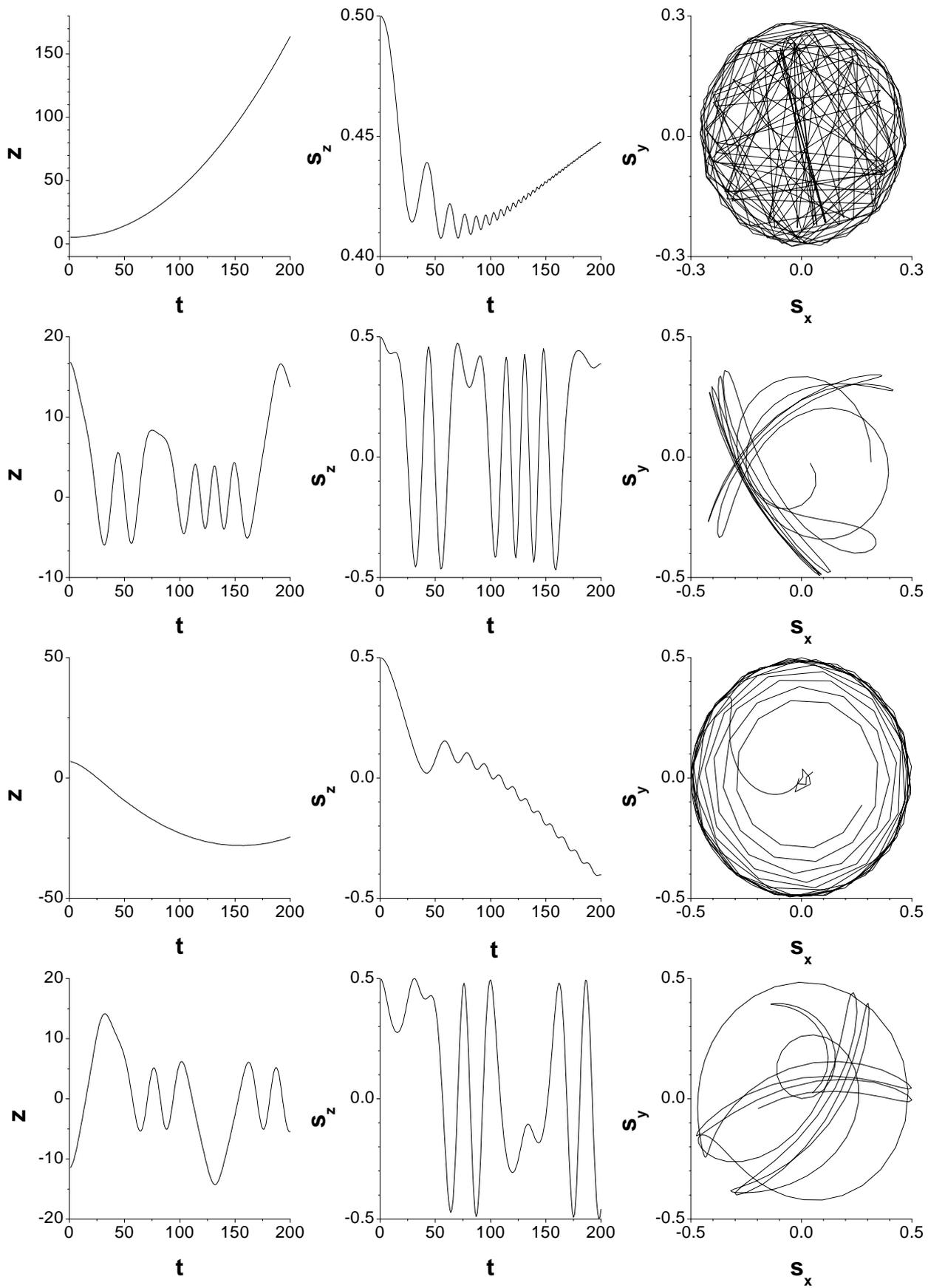
\begin{figure}[tbp]
\caption{Typical time dependence of the z-th coordinate of the particle, the
z-th projection of its spin and the projection of its spin in the x-y plane.
Small changes in the initial conditions result in entirely different time
dependence of these parameters.}
\label{fig4}
\end{figure}


\begin{references}
\bibitem{tomonaga}  For the historic account of the spin see: S.-I.
Tomonaga, {\it The story of spin, }The University of Chicago Press, Chicago
(1997)

\bibitem{dirac}  P. A. M. Dirac, Proc. Roy. Soc. London {\bf A117}, 610
(1928)

\bibitem{mcgregor}  Interesting ideas can be developed around the classical
concept of the spin, as extensively discussed in: M. H. MacGregor, {\it The
Nature of the elementary particle, }Springer (1978), Lecture Notes in
Physics 81

\bibitem{berezin}  The meaning of the term {\it classical }is often
difficult to comprehend, as for example in: F. A. Berezin and M. S. Marinov,
Ann. Phys. (N.Y.) {\bf 104}, 336 (1977)

\bibitem{bohr}  N. Bohr, J. Chem. Soc. {\bf 134}, 349 (1932)

\bibitem{mott}  N. F. Mott and H. S. W. Massey, {\it The theory of atomic
collisions,} Claredon Press, Oxford (1965), pp.214-219

\bibitem{bargmann}  V. Bargmann, L. Michel and V. Telegdi, Phys. Rev. Lett. 
{\bf 2}, 435 (1959)

\bibitem{barut}  A. O. Barut, {\it Electrodynamics and classical theory of
fields and particles, }The MacMillan (new York 1964), pp.73

\bibitem{batelaan}  H. Batelaan, T. J. Gay and J. J. Schwendiman, Phys. Rev.
Lett. {\bf 79}, 4517 (1997)

\bibitem{garraway}  B. M. Garraway and S. Stenholm,

\bibitem{bos1}  S. D. Bosanac, {\it Classical Dynamics with the Uncertainty
Principle}, in {\it From Simplicity to Complexity: Information, Interaction,
Emergence}, Edited by: A. Mueller, K. Mainzer and W. Saltzer, Vieweg-Verlag,
Wiesbaden (1997)

\bibitem{bos2}  H. Skenderovi\'{c} and S. D. Bosanac, Zeit. f. Phys. {\bf D35%
}, 107 (1995)

\bibitem{bos3}  N. Do\v{s}li\'{c} and S. D. Bosanac, Mol. Phys. {\bf A90},
599 (1997)

\bibitem{bos4}  S. D. Bosanac, Physica Scripta {\bf 57}, 171 (1998)

\bibitem{bos5}  N. Klipa and S. D. Bosanac, Int. J. Theor. Phys., Group
Theory and Nonlinear Optics {\bf 7}, 15, (2000)
(http://arXiv.org/abs/quant-ph/0010089)

\bibitem{bos6}  S. D. Bosanac, J. Math. Phys. {\bf 38}, 3895, (1997)

\bibitem{wigner}  E. Wigner, Phys. Rev. {\bf 40}, 749 (1932)

\bibitem{hillery}  M. Hillery, R. F. O'Connel, M. O. Scully and E. P.
Wigner, Phys. Rep. {\bf 106}, 122 (1984)

\bibitem{carruthers}  P. Carruthers and F. Zachariasen, Rev. Mod. Phys. {\bf %
55}, 245 (1983)

\bibitem{moyal}  J. E. Moyal, Proc. Camb. Phil. Soc. {\bf 45}, 99, (1949)

\bibitem{gold}  H. Goldstein, {\it Classical Mechanics}, Addison-Wesley
(1981), pp. 148
\end{references}
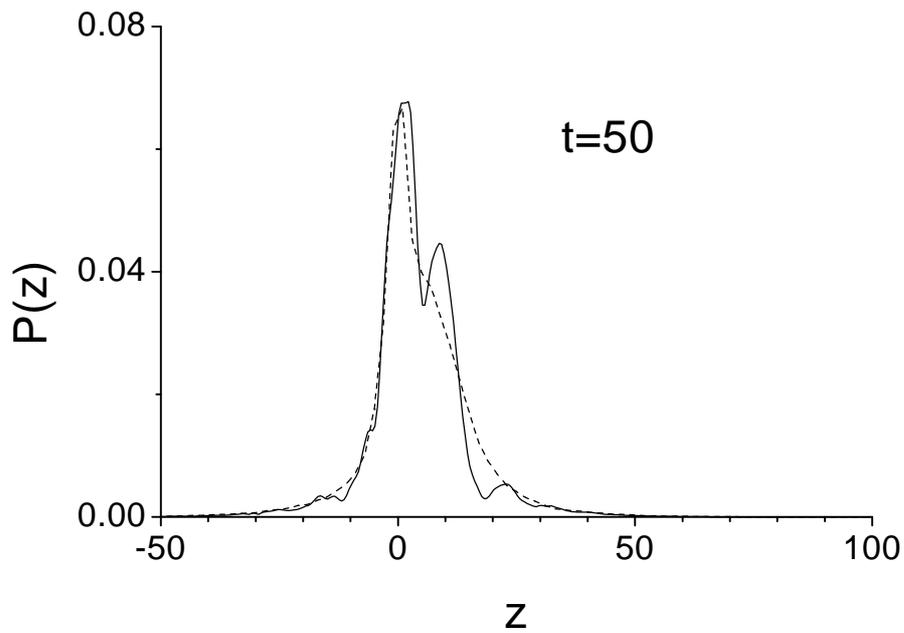
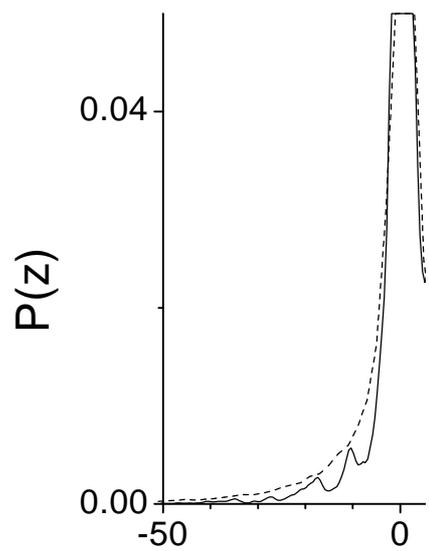
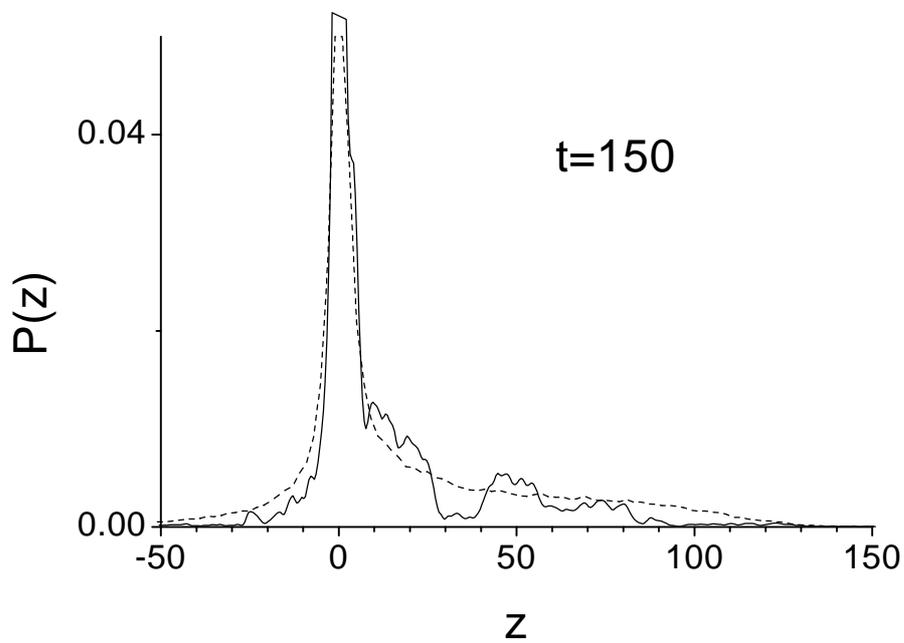
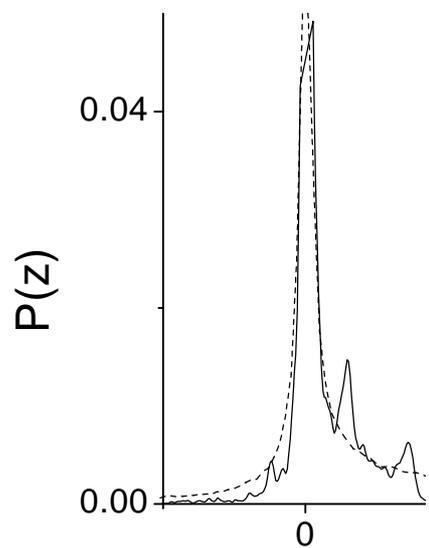
\end{document}